\pdfoutput=1

\documentclass[conference]{IEEEtran}

\ifCLASSINFOpdf
\else
\fi

\usepackage{graphicx}
\def\lf{\left\lfloor}   
\def\rf{\right\rfloor}
\usepackage{booktabs}
\usepackage{array}
\usepackage{amsmath}
\usepackage{mathtools}
\usepackage{cite}
\usepackage{verbatim}
\usepackage{tikzpagenodes}

\makeatletter
\renewcommand*\env@matrix[1][c]{\hskip -\arraycolsep
  \let\@ifnextchar\new@ifnextchar
  \array{*\c@MaxMatrixCols #1}}
\makeatother
\makeatletter
\newcommand*{\rom}[1]{\expandafter\@slowromancap\romannumeral #1@}
\makeatother

\begin{document}

\title{Towards Design Space Exploration and Optimization of Fast Algorithms for Convolutional Neural Networks (CNNs) on FPGAs}
\author{\IEEEauthorblockN{Afzal Ahmad, Muhammad Adeel Pasha}
\IEEEauthorblockA{Department of Electrical Engineering, Syed Babar Ali School of Science and Engineering (SBASSE),\\
Lahore University of Management Sciences (LUMS), Lahore, Pakistan\\
Email: {afzal.ahmad@lums.edu.pk, adeel.pasha@lums.edu.pk}
}
}
\maketitle

\begin{tikzpicture}[remember picture,overlay]
    \node[align=center,text=black] at ([yshift=1em]current page text area.north) {Preprint: Accepted at 22nd IEEE Design, Automation \& Test in Europe Conference \& Exhibition (DATE'19)};
\end{tikzpicture}%

\begin{abstract}
Convolutional Neural Networks (CNNs) have gained widespread popularity in the field of computer vision and image processing. Due to huge computational requirements of CNNs, dedicated hardware-based implementations are being explored to improve their performance. Hardware platforms such as Field Programmable Gate Arrays (FPGAs) are widely being used to design parallel architectures for this purpose. In this paper, we analyze Winograd minimal filtering or fast convolution algorithms to reduce the arithmetic complexity of convolutional layers of CNNs. We explore a complex design space to find the sets of parameters that result in improved throughput and power-efficiency. We also design a pipelined and parallel Winograd convolution engine that improves the throughput and power-efficiency while reducing the computational complexity of the overall system. Our proposed designs show up to 4.75$\times$ and 1.44$\times$ improvements in throughput and power-efficiency, respectively, in comparison to the state-of-the-art design while using approximately 2.67$\times$ more multipliers. Furthermore, we obtain savings of up to 53.6\% in logic resources compared with the state-of-the-art implementation.
\end{abstract}

\IEEEpeerreviewmaketitle

\section{Introduction}
Convolutional Neural Networks (CNNs) have widely been used for object detection problems in image processing and computer vision domains. Although CNNs are known to give unprecedented results on these problems, they are limited by their copious computational requirements. Extensive work is being done on acceleration of these networks while reducing the computation cost. Graphics Processing Units (GPUs) are widely being used for this purpose owing to their immense parallelization capabilities~\cite{cudaAccel, alexNet}. While general-purpose GPUs give great performance results for CNNs, dedicated hardware-based approaches are needed for systems that have stringent time and accuracy constraints e.g. self-driving cars, autonomous robots and unmanned aerial vehicles. 

Field Programmable Gate Array (FPGA) based accelerators for neural networks have been proposed aiming at increasing the parallelism of architectures while minimizing the latency and improving the throughput~\cite{podili2017fast, endtoend, rahman2016efficient}. Optimization techniques aiming at reducing the arithmetic computational complexities have also been proposed and have shown to cut the cost of forward pass of neural networks by major factors. Fast Fourier Transform (FFT) based acceleration schemes have shown to reduce the computation complexities of convolutions but are only feasible for large kernel sizes~\cite{fbfft} whereas modern state-of-the-art CNN architectures mostly involve smaller kernels~\cite{resnets, vgg16} making FFT-based designs less favorable. 

Hence, fast convolution algorithms based on Winograd minimal filtering~\cite{winograd1980} have been proposed and show prominent performance gains for smaller kernel sizes. These algorithms are aimed at reducing the arithmetic complexity of convolution operations by reducing the number of expensive operations, e.g. multiplications, at the cost of cheaper operations, e.g. additions; an optimization known as algorithmic strength reduction used in filter design.

In this paper, we design hardware accelerators for CNNs based on Winograd minimal filtering algorithms. We also propose optimization schemes for reducing the arithmetic and logic resource costs of transforms. Meanwhile, there is a design space that needs to be explored to find the variation in performance and hardware utilization given a set of parameters associated with the minimal filtering algorithms. The major contributions of this paper are as follows:
\begin{itemize}
\item We performed a comprehensive design space exploration to find the optimal parameters of Winograd minimal filtering algorithms that give the most benefits in throughput and resource efficiency.
\item We improved the hardware design of the state-of-the-art one-dimensional (1D) Winograd convolution engine~\cite{podili2017fast}. Our proposed design results in about 53.6\% logic resource reduction and improves the power-efficiency by about 1.44$\times$ over the reported design.
\item We designed systems implementing Winograd minimal filtering algorithms of higher order, improving throughput and resource efficiency by 4.75$\times$ and 1.78$\times$ over the previously reported design.
\end{itemize}

The rest of the paper is structured as follows. In Section\,II, we discuss the basics of CNNs and present some related work. In Section\,III, we present a design space exploration to quantify the impact of varying different fast convolution parameters on resultant arithmetic complexity. Section\,IV contains details of our proposed hardware implementation of Winograd convolution engine. In Section\,V, we present the synthesis results of our proposed FPGA implementations and compare their performance, resource utilization and power-efficiency with the state-of-the-art designs. The paper is then concluded in Section\,VI.

\section{Convolutional Neural Networks (CNNs)}
CNNs are a special class of deep neural networks used for object detection in images. CNNs consist of two types of layers: (i) convolutional layers and (ii) fully-connected layers. A convolutional layer takes an input feature map of $N$ minibatch images, each with height, width and number of channels, $H$, $W$ and $C$, respectively. A filter or kernel of size $r \times r$ pixels having $C$ channels is also provided as input to the convolutional layer and a two-dimensional (2D) output feature map is generated. $K$ such kernels are applied to the input feature map to generate a three-dimensional (3D) output feature map. Convolutional layers extract prominent features from the input feature maps. They have been shown to consume more than 90\% of the total execution time of modern CNNs~\cite{cong2014minimizing}. Hence, acceleration of convolutional layers is the key to improving the performance of these deep neural networks. Below, we briefly introduce two major approaches for performing convolutions in CNNs.

\subsection{Spatial Convolution}
A spatial convolution operation is multiplication and accumulation of corresponding elements of an input feature map and a kernel to generate a single output pixel. The kernel is then swept across the input feature map to generate a single channel of the output feature map. Denoting a kernel map as $G_{k,c,u,v}$ and a single tile of input feature map as $D_{i,c,x,y}$, an output pixel $Y_{i,k,x,y}$ is calculated using a spatial convolution algorithm~\cite{alavin:1} as,

\begin{equation}
Y_{i,k,x,y}=\sum_{c=1}^{C}\sum_{v=1}^{r}\sum_{u=1}^{r}D_{i,c,x+u,y+v}\ G_{k,c,u,v}
\end{equation}

Where $x$ and $y$ are coordinates of the feature map tile, $i$ is image number in the batch, $u$ and $v$ are iterators over kernel while $c$ is the iterator over channels and $k$ is kernel index.

\subsection{Winograd Minimal Filtering Algorithms}
Winograd minimal filtering or fast convolutions~\cite{alavin:1} are an application of algorithmic strength reduction whereby more complex operations, e.g. multiplications, are replaced by equivalent representations in less complex operations, e.g. additions. This effectively increases the efficiency with which convolutional layers can be computed. A one-dimensional (1D) minimal algorithm represented as $F(m,r)$ needs $m+r-1$ multiplications to compute $m$ outputs using a filter size $r$. This algorithm takes $m+r-1$ inputs, $r$ filter elements and implements the matrix equation~\cite{alavin:1},

\begin{equation}
\vspace{-2mm}
Y=A^T[(B^Td)\odot\,(Gg)]
\label{Winograd1D}
\end{equation}

Where $A$, $B$ and $G$ are constant matrices for the inverse, data and filter transforms while $d$ and $g$ are data and filter inputs, respectively. This algorithm only needs $m+r-1$ multiplications for convolution as opposed to $m \times r$ multiplications needed by the spatial convolution. However, this comes at a cost of additional matrix transformations that are composed of additions, subtractions and constant multiplications which are relatively cheaper operations. 

A 2D minimal filtering algorithm $F(m \times m,r \times r)$ can be obtained by nesting a 1D minimal algorithm within itself. A 2D minimal algorithm takes a 2D image tile of size $(m+r-1) \times (m+r-1)$ and an $r \times r$ kernel to generate an $m \times m$ output tile. This algorithm only needs $(m+r-1)^2$ multiplications as opposed to $m^2 \times r^2$ needed by spatial convolution. A 2D minimal algorithm can be written as,

\begin{equation}
\vspace{-2mm}
Y=A^T[U\odot V]A
\label{Winograd2D}
\end{equation}

Where $U=B^TdB$ is the data transform, $V=GgG^T$ is the filter transform and $Y=A^TMA$ is the inverse transform where $M=U\odot V$ is the element-wise multiplication stage. The authors in~\cite{alavin:1} computed transformation matrices $B$, $G$ and $A$ for several different configurations of $m$ and $r$.

For parameters $m,r$ of 2D minimal algorithms, $d$ is the input data tile of size $(m+r-1) \times (m+r-1)$ while $g$ is the kernel matrix of size $r \times r$. Complexity analysis of Winograd filtering relative to spatial convolutions is done in Section\,III.

\subsection{Previous Work}

Ref.~\cite{alavin:1} first studied GPU implementations of fast algorithms based on Winograd minimal filtering, achieving a speedup of up to 7.28$\times$ over NVIDIA CUDA Deep Neural Networks Library (cuDNN). cuDNN now incorporates Winograd filtering based convolutions. FFTs based computations of convolutions is studied in~\cite{fbfft} but show savings only for high kernel sizes and are not applicable to most layers of modern CNNs where the trend is towards smaller kernel sizes. In hardware implementations,~\cite{endtoend} studied a fully-pipelined FPGA accelerator implementing techniques for both convolutional and fully-connected layers. Ref.~\cite{rahman2016efficient} proposed a CNN accelerator using 3D arrays of Multiply-and-Accumulate (MAC) units and a complex data reuse network to maximize the resource usage and throughput while~\cite{podili2017fast} explored a highly parallel, scalable and pipelined convolution engine implementing Winograd minimal filtering using fixed parameters on an Altera Stratix\,V\,GT FPGA.

In the next section, we explore the effects of increasing the output tile size, $m$, on the multiplication complexity of the element-wise multiplication stage, $M$, and the arithmetic complexity of the transforms' stages, $U$, $V$ and $Y$.

\section{Design Space Exploration}

We performed an extensive design space exploration with regards to the arithmetic complexity and performance of Winograd minimal algorithms using VGG16 network D~\cite{vgg16} as the CNN model of choice. This CNN architecture uses kernels with sizes $3 \times 3$ pixels in all of its layers, hence the same convolution engine can be applied to all the convolutional layers of this CNN model. The design space shows a tradeoff between the decrease in multiplication complexity of element-wise multiplication stage $U \odot V$ and the increase in arithmetic complexity of data, filter and inverse transform stages. Similarly, it shows an increase in throughput with an increase in output tile size, $m$, of minimal filtering functions.
\vspace{-1mm}
\subsection{Multiplication Complexity}

The multiplication complexity of the element-wise multiplication stage is given by,

\begin{equation}
O_m= \frac{NHWCK}{m^2}(m+r-1)^2
\label{MComplex}
\end{equation}

\begin{figure}[t]
	\centering
  \includegraphics[width=3.48in]{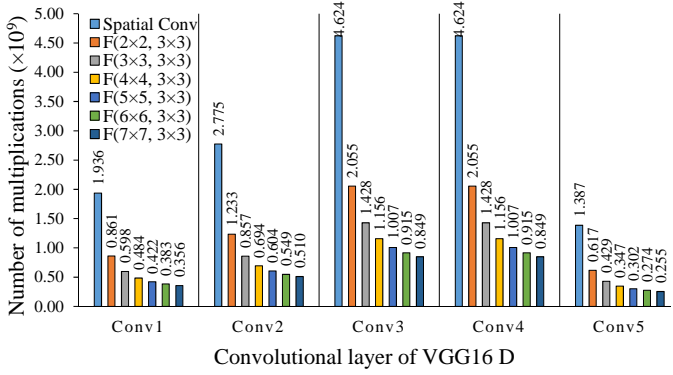}
	\vspace{-6mm}
	\caption{Decrease in multiplication complexity $O_m$ with $m$}
	\label{mulComp}
\end{figure}

Where for spatial convolutions, $m=1$. The arithmetic complexity of the multiplication stage hence decreases quadratically with an increase in output tile size $m$. Fig.~\ref{mulComp} shows the decrease in multiplication complexity with an increase in the order of Winograd convolution function $m$ for the different group layers of VGG16 network D. Higher-order minimal filtering functions do reduce the multiplication complexity but with diminishing gains and also come at a cost of increased complexity of the data, filter and inverse transform stages. Furthermore, since each Processing Element (PE) performs one multiplication per input, the number of multipliers required per PE increase with increase in input tile size $(m+r-1)^2$. So for a fixed set of hardware resources, the number of parallel instantiated PEs decreases with the increase in $m$. 

\subsection{Transforms' Complexities}

Algorithmic strength reduction decreases the complexity of expensive operations, e.g. multiplications, for an increase in cheaper operations, e.g. additions and constant multiplications. Data, filter and inverse transforms are all exclusively composed of these cheaper operations. The arithmetic complexities of these transforms are given by~\cite{alavin:1},

\begin{equation}
 \left.\begin{aligned}
        T(D)&=\frac{\beta}{m^2}NHWC,\\
        T(F)&=\gamma CK,\\
				T(I)&=\frac{\delta}{m^2}NHWK.
       \end{aligned}
 \right\}
 \qquad
	\label{TComplex}
\end{equation}

Where $\beta$, $\gamma$ and $\delta$ are number of floating point operations required by the data, filter and inverse transforms of single tiles, respectively.

\begin{figure}[t!]
	\centering
  \includegraphics[width=3.48in]{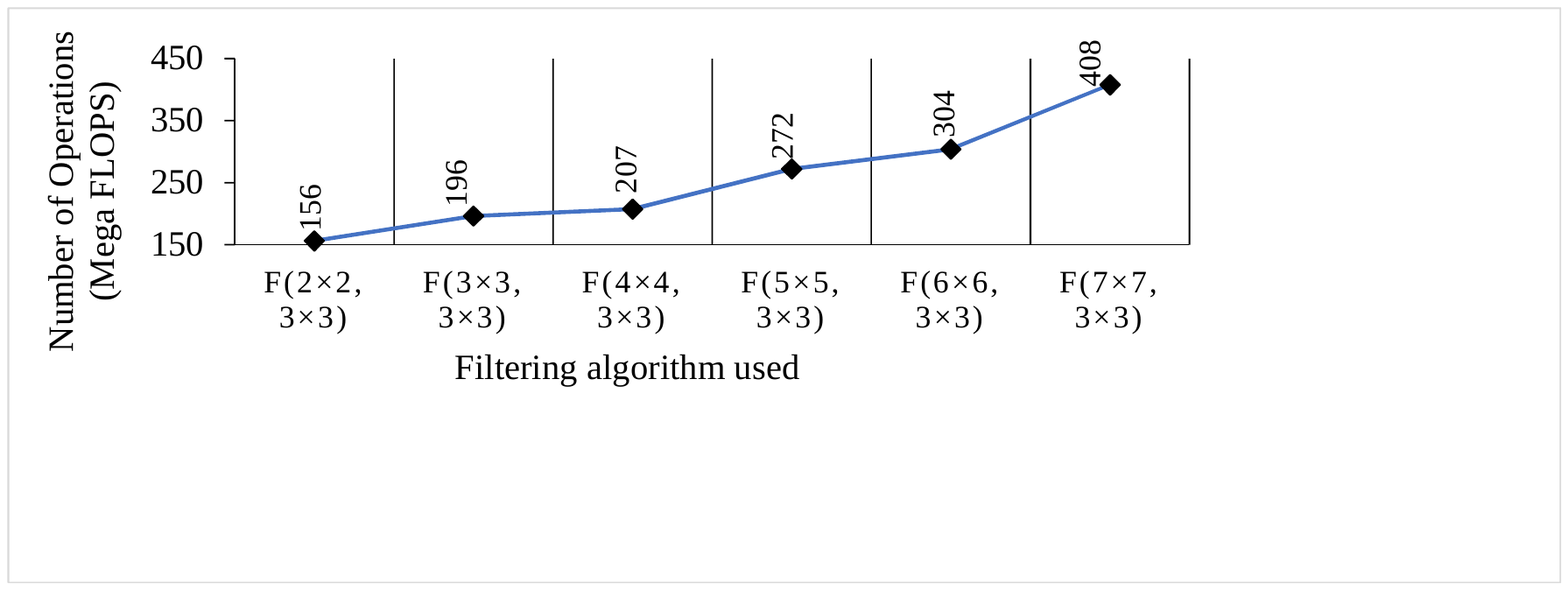}
	\vspace{-6mm}
	\caption{Increase in net transform complexity $O_t$ with $m$}
	\label{TransformsCurve}
\end{figure}

The net arithmetic complexity of the transforms, $O_t$ is simply the sum of the arithmetic complexities of the three transform stages,
\begin{equation}O_t= T(D)+T(F)+T(I) \label{TotComplex}\end{equation}
Fig.~\ref{TransformsCurve} shows the quadratic increase in arithmetic complexity of transforms with the increase in output tile size $m$ for VGG16\,D. This increase can be attributed to the increase in complexities of the individual transform stages; $\beta$, $\gamma$ and $\delta$ with increase in $m$.

\begin{figure}[t!]
	\centering
  \includegraphics[width=3.48in]{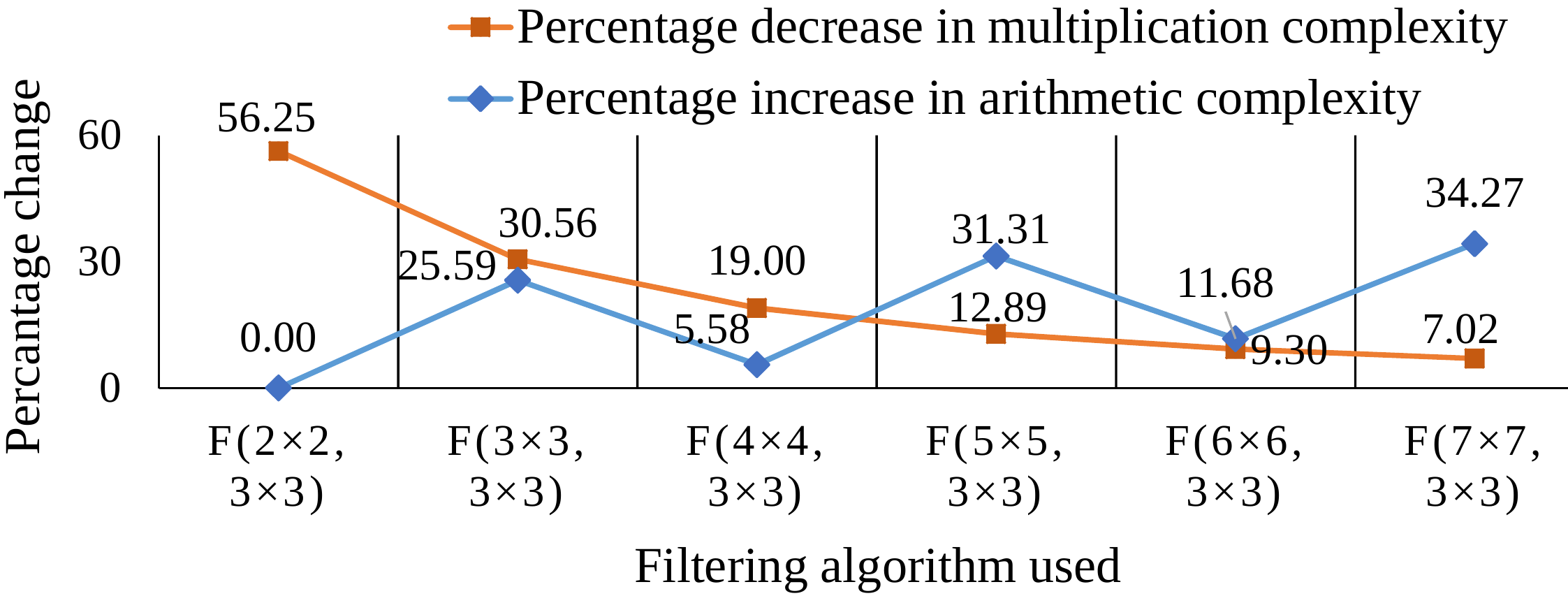}
	\vspace{-6mm}
	\caption{Percentage variations of complexities with $m$}
	\label{percentageChangeCurves}
\end{figure}

\subsection{Discussion} \label{dse}

\begin{figure*}[t]
	\centering
  \includegraphics[width=14.0cm, height=6cm]{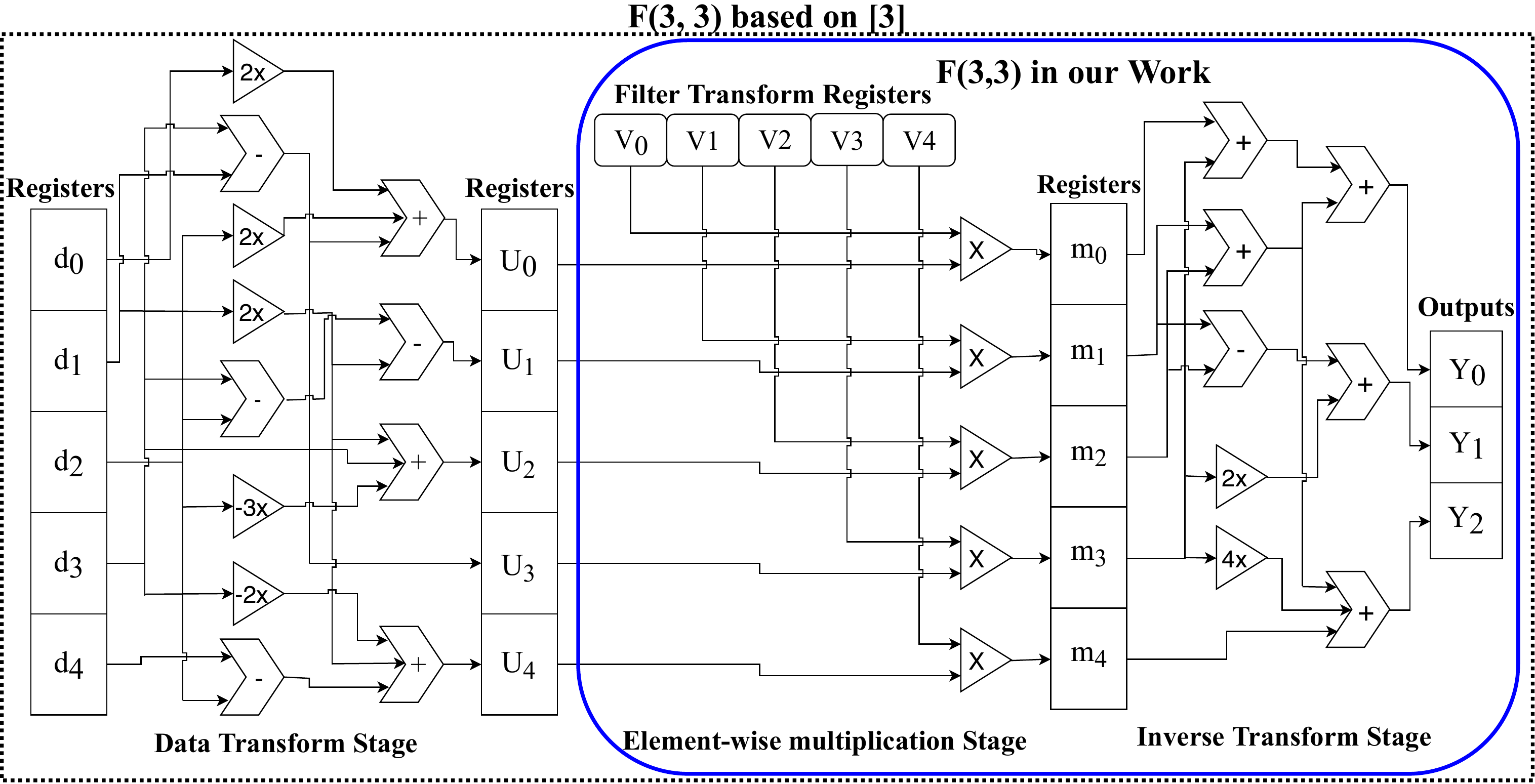}
	\vspace{-3mm}
	\caption{Comparison of our 1D convolution engine, $F(3, 3)$, with that of \cite{podili2017fast}}
	\label{Engine1D}
\vspace{-5mm}
\end{figure*}

While the multiplication complexity decreases quadratically with the increase in $m$, the arithmetic complexity of the transforms sees an overall quadratic increase. Hence, it results in decreasing savings from using minimal filtering algorithms for performing the convolutions for larger output tile sizes $m$. Fig.~\ref{percentageChangeCurves} shows that going from $m=3$ to $4$, the relative multiplication complexity decrease is about 19\% while the relative transform complexity increases by only about 5.58\%. While this case is favorable, when going higher to $m=5$, the relative decrease in multiplication complexity is only 12.89\% while the relative transforms complexity increases by 31.31\%, making the overall design less favorable than $m=4$. Hence, for $m\geq5$, the increase in the overhead of transforms complexity has outweighed the savings from the decrease in multiplication complexity making it infeasible to perform Winograd filtering based convolution. The decrease in multiplication complexity directly translates to an increase in throughput while the increase in arithmetic complexity translates to an increase in logic resources required to implement the design and resultant power dissipation.

\section{Proposed Hardware Design}

We implemented pipelined hardware architectures for 3 different configurations of 2D Winograd minimal filtering algorithms; $F(m \times m,r \times r)$ where the kernel size is constant, $r=3$ and output tile sizes, $m=2,3,4$ which show high throughput gain while incurring bearable loss due to increase in transforms complexity as discussed in previous section. Due to space limitation, we will only discuss the design of $F(3 \times 3,3 \times 3)$ but we will present the implementation results for all three designs in Section~\ref{results}. We used single precision floats without any quantization scheme for the sake of simplicity and high precision.

\subsection{2D Winograd Convolution Engine}

A single PE module of $F(3 \times 3,3 \times 3)$ takes input tiles of sizes $(m+r-1)^2=5^2=25$ while using the same number of multipliers to implement the 2D minimal algorithm (\ref{Winograd2D}). The computation of each $m \times m=3 \times 3$ tile of output is divided into 3 distinct stages: (i) data transform stage, (ii) element-wise multiplication stage and (iii) inverse transform stage. Filter transforms are only dependent on the kernel matrices and hence are assumed to be precomputed. The three stages are pipelined to optimize the throughput. 

\begin{figure}[t]
	\centering
  \includegraphics[width=3.48in]{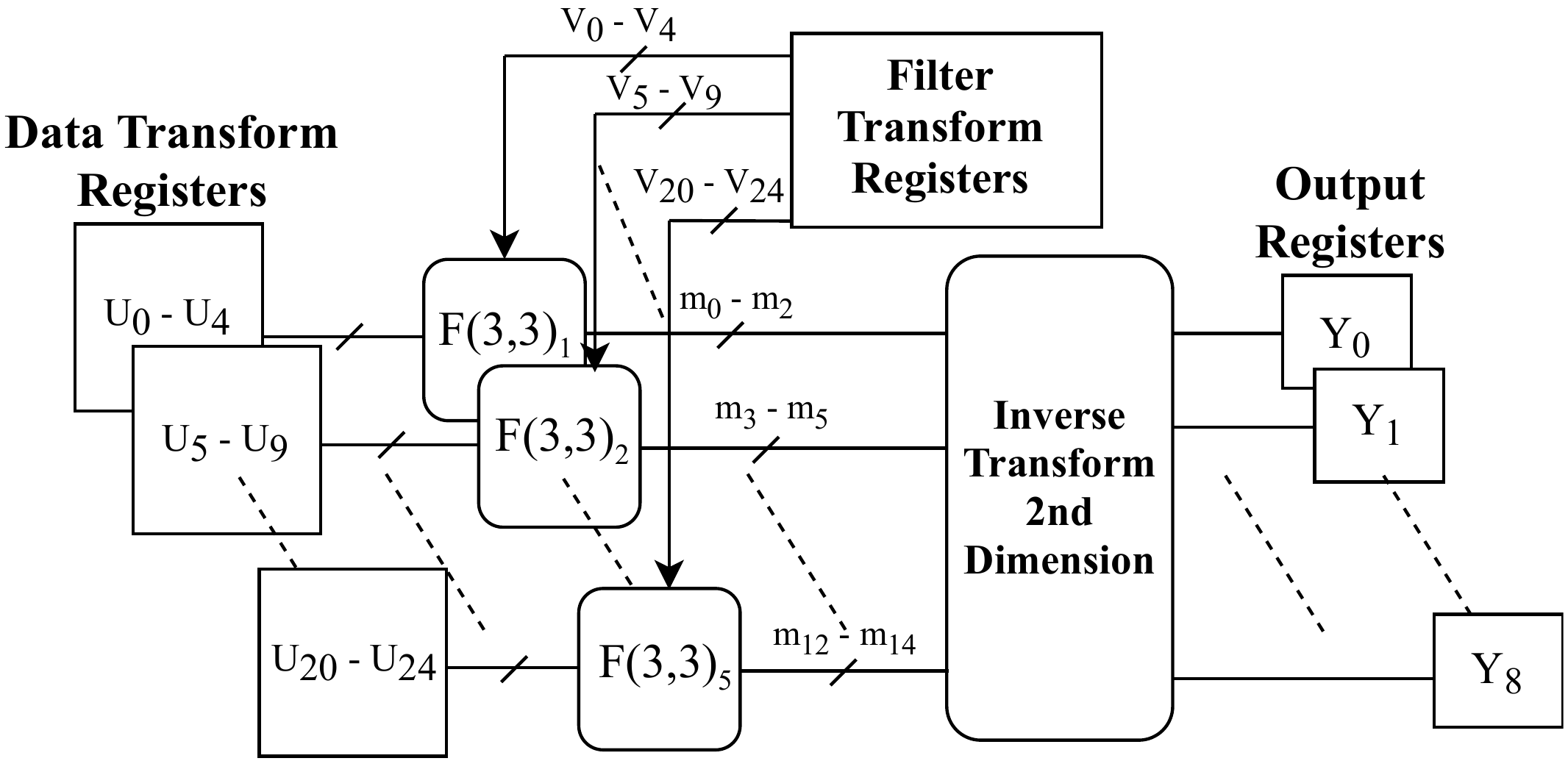}
	\vspace{-6mm}
	\caption{Layout for a PE, $F(3\times3, 3\times3)$}
	\label{PE}
	\vspace{-4mm}
\end{figure}

Fig.~\ref{Engine1D} shows the layout of our 1D Winograd convolution engine $F(3,3)$ in solid blue box. This convolution engine takes $(3+3-1)=5$ units of data transformed image tile, $U$, and filter transformed kernels, $V$ as inputs and performs the element-wise multiplication and the inverse transform operation in 1D. The 1D Winograd convolution engine is nested within itself to form 2D Winograd convolution engine shown in Fig.~\ref{PE}. The 2D convolution engine takes $(3+3-1)^2=25$ units of $U$ and $V$ and implements the 2D minimal algorithm according to (\ref{Winograd2D}). Due to pipelining, the throughput achieved is $m^2=9$ units of outputs per clock cycle per PE as opposed to the work in~\cite{podili2017fast} where $4$ units of outputs are computed per clock cycle per PE. Hence, we achieve $\frac{9}{4}=2.25\times$ higher throughput than the previous work while using only $\frac{25}{16}=1.56\times$ more number of multipliers per PE. Throughput is linearly related to the number of PEs employed in the design since each parallel PE gives 9 units of output per clock cycle.

\subsection{System-Level Description}

\begin{figure}[t]
  \includegraphics[width=3.49in]{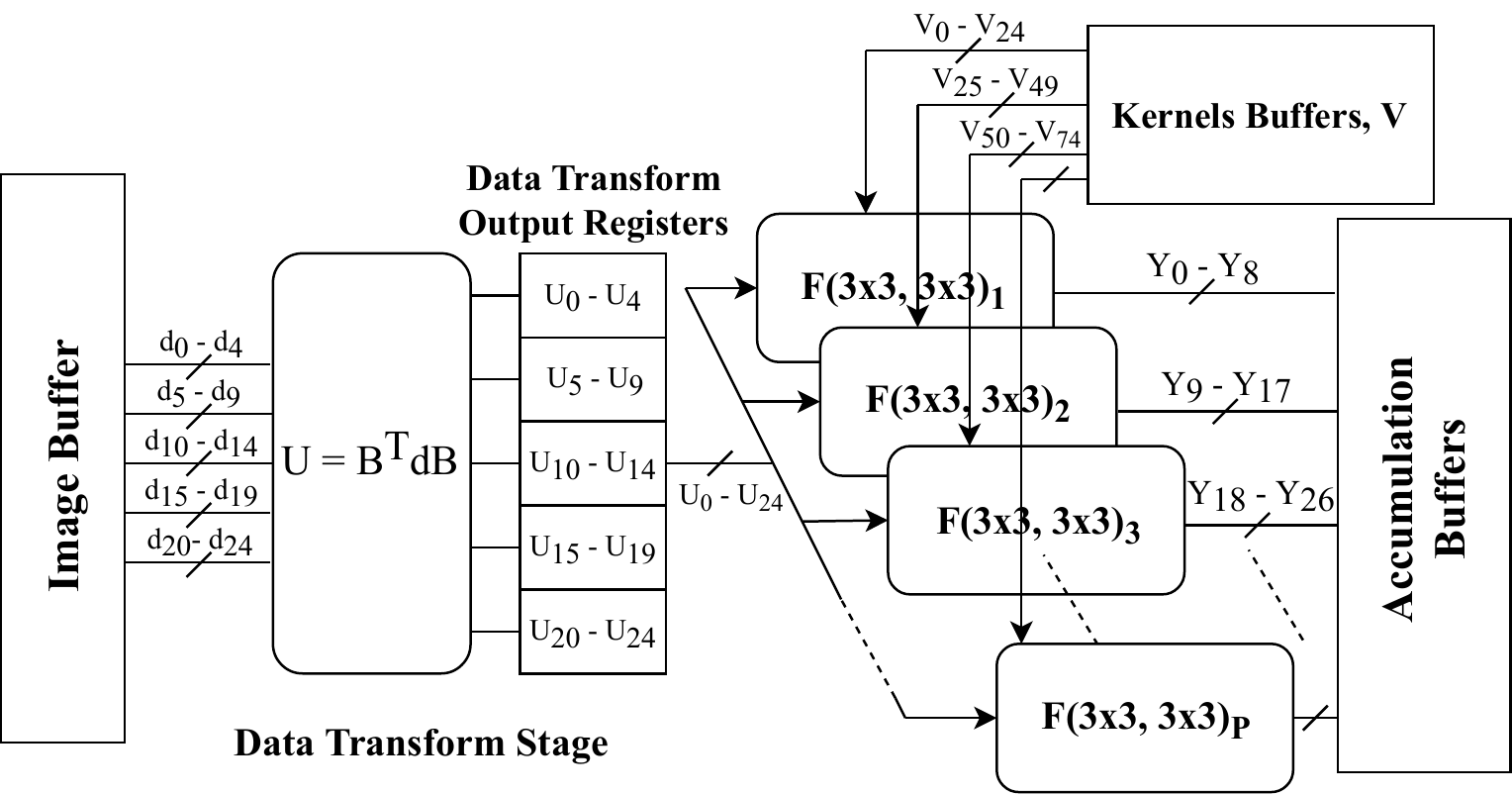}
	\vspace{-6mm}
	\caption{System level implementation}
	\label{SystemLvl}
	\vspace{-1mm}
\end{figure}

Each PE takes precomputed $U$ and $V$ as inputs and performs the 2D minimal algorithm to compute the output $Y$. Although $V$ can be precomputed even before running a forward pass of the CNN, $U$ needs to be computed in the datapath of convolution engine since it is dependent on the input image data. 
Fig.~\ref{SystemLvl} shows the system-level implementation of our design. In each clock cycle, a $5 \times 5$ tile of image data is passed through the data transform stage which is composed of simple arithmetic and constant multiplications that can easily be implemented using shifters and adders. The same resultant data transformed input tile $U$ is then forwarded to $P$ parallel PEs. While all $P$ PEs receive the same data transformed input tile, $U$, each PE receives its own filter transformed kernel $V$. Consequently, in a single clock cycle, $P$ different kernel tiles are applied to the data transformed input tile $U$. The outputs of the convolution between the input tile and each kernel tile is then stored in a buffer at the output and then accumulated over the next $C$ clock cycles to compute the result of the convolution across the 3rd dimension i.e. channels. 

\subsection{Transform Complexity of Implementation}

Although the net arithmetic complexity of transforms increase with the increase in output tile size $m$, we can precompute the filter transforms; hence reducing the overall increase in arithmetic complexity of our design to the sum of arithmetic complexities of data and inverse transforms only. The overall transforms complexity of our design, $O_T$ is given by,

\begin{equation}O_T=\frac{NHWCK}{m^2}(\frac{\beta}{P} + \delta)\label{ImplTotComplex}\end{equation}

Where $P$ is number of parallel PEs used in the design given by,

\begin{equation}P=\lf \frac{m_T}{m_P}\rf=\lf \frac{m_T}{(m+r-1)^2}\rf \label{PEq}\end{equation}

Where $m_T$ and $m_P$ are the total number of available multipliers and number of multipliers per PE, respectively.
Making the data transform stage independent of the PEs allows us to divide the arithmetic complexity of the data transform by the factor $P$, the number of parallel PEs. For $F(2 \times 2,3 \times 3)$ using 16 parallel PEs, the increase in transform complexity of our design relative to spatial convolutions is only 1.5$\times$ while for the state-of-the-art design~\cite{podili2017fast}, this increase is 2.33$\times$.

\subsection{Throughput Calculations}

\begin{figure}[t]
	\centering
  \includegraphics[width=3.49in]{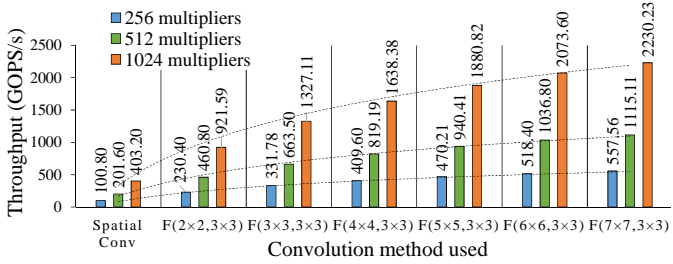}
	\vspace{-6mm}
	\caption{Throughput variation with $m$ and number of multipliers}
	\label{throughputCurve}
\end{figure}

The total time required to calculate the output feature map from an input feature map is given by~\cite{podili2017fast},

\begin{equation}T_t= (\frac{NHWCK}{m^2P}+D_p-1)t_c \label{ThroughputEq}\end{equation}

Where $P$ is the number of parallel PEs, $D_p$ is the pipeline depth and $t_c$ is the clock cycle time.

Two observations can be made from Eqs.~(\ref{PEq}) and (\ref{ThroughputEq})

\begin{enumerate}
\item For a fixed number of multipliers, $m_T$, increasing the factor $m$ decreases the parallelism factor $P$ according to (\ref{PEq}) since the design needs more multipliers per PE for larger output tile sizes. The overall effect of increasing $m$ is a quadratically decreasing $T_t$.
\item For a fixed set of minimal filtering function parameters $r$ and $m$, increasing the number of available multipliers, $m_T$, leads to a linear increase in $P$ according to (\ref{PEq}) which, in turn, leads to a linearly decreasing $T_t$.
\end{enumerate}

Overall throughput of the system is given by,

\begin{equation}Throughput= \frac{O_S}{T_t}\end{equation}

Where $O_S$ is the arithmetic complexity of spatial convolution. Fig.~\ref{throughputCurve} shows the variation of throughput with output tile size $m$ and number of multipliers available for a system with an operating frequency of 200\,MHz. Throughput increases linearly with the number of multipliers used in the system and quadratically with increasing output tile size $m$.

\subsection{Comparison with the State-of-the-Art Implementation}

Ref.~\cite{podili2017fast} used a similar pipelined structure to implement 2D minimal function $F(2\times\,2,3\times\,3)$. Fig.~\ref{Engine1D} shows our 1D Winograd convolution engine for $F(3,3)$ in solid blue box while that based on~\cite{podili2017fast} in dotted black. Ref.~\cite{podili2017fast} implemented the data transform stage as a part of the 1D convolution engine which computes the data transform independently for each PE. This is redundant and wasteful since all PEs receive the same data transformed input. We separated the data transform stage and computed it only once for all $P$ PEs. Consequently, we amortized the cost of data transform over the number of parallel PEs saving us hardware logic resources required to implement the designs. Our implementation performs the data transforms only once per clock cycle instead of $P$ times per clock cycle as done in~\cite{podili2017fast}. 

As a second contribution, we implemented hardware designs for higher order Winograd filtering functions leading to higher throughput compared with the reported works. Although we only discussed the design for $(m,r=3,3)$ due to lack of space, our higher order design of $(m,r=4,3)$ gives a higher throughput but is also more complex having dense transform stages. Hence, even though $F(4 \times 4,3 \times 3)$ uses more multipliers per PE allowing us to only instantiate 19\,PEs in parallel, we get the highest performance that is consistent with the findings discussed in Section~\ref{dse}.

\section{Experimental Results} \label{results}
 
In this section, we present the synthesis results for our convolution engines implementing $F(m\times m,r\times r)$ where $m=2,3,4$ and $r=3$ on a Xilinx Virtex\,7 FPGA. We will also compare our results with the state-of-the-art implementations~\cite{podili2017fast, qiu2016going}.

\subsection{Resource Utilization}
For Winograd minimal filtering parameters $m,r=4,3$, we were able to design the system implementing Winograd filtering with a maximum of 19 parallel PEs since beyond this point, more PEs could not be instantiated with the remaining DSPs on the target FPGA. Due to better data reuse of the data transform stage (our first contribution), we were able to reduce the percentage utilization of LUTs by approximately 53.6\% for 19 parallel PEs as shown in Table~\ref{tab:Table1}. We achieved higher savings in slice logic utilization for high number of parallel PEs. For an implementation based on the reference design~\cite{podili2017fast}, the required number of slice LUTs increases by about 12224 LUTs per PE while for our implementation, this increase is only about 5312 LUTs per PE. This reduction in slice logic directly translates to a more power-efficient design as discussed in the next subsection. 

\subsection{Performance Comparison}

\begin{table}[t]
  \begin{center}
		\caption{Resource utilization for 19\,PEs $F(4\times 4,\ 3\times 3)$}
		\vspace{-3mm}
		\label{tab:Table1}
		\begin{tabular}{|l|c|c|c|c|}
		\hline
      \textbf{} & Registers & LUTs & DSPs & Multipliers\\	\hline
      Design based on~\cite{podili2017fast} & 97052 & 232256 & 2736 & 684\\ \hline
      Our proposed design & 76500 &  107839 & 2736 & 684\\ \hline
      Available resources & 607200 & 303600 & 2800 & 700\\ \hline
    \end{tabular}
  \end{center}
	\vspace{-2mm}
\end{table}

In this subsection, we compare the performance results of our designs with previous FPGA based implementations of CNNs~\cite{podili2017fast, qiu2016going}. We performed latency and throughput calculations assuming that double buffering is employed at both image and kernel buffers and enough memory bandwidth is available to refill both buffers without having to wait for more input data to be available. 

Table~\ref{tab:Table2} shows that our design of Winograd convolution engine for $m=2$ gives the same latency and hence throughput as~\cite{podili2017fast} when using the same number of multipliers. This is because latency of the system is dependent on the element-wise multiplication stage which is the slowest pipeline stage in the two designs and moving the data transform stage out of individual PEs has no effect on this stage. However, this modification to the data transform stage does decrease the slice logic utilization and leads to an increase in power-efficiency. For $m=2$, we achieve 1.44$\times$ and 2.12$\times$ better power-efficiency than~\cite{podili2017fast} and~\cite{qiu2016going}, respectively. For $m=4$, we achieve approximately the same power-efficiency as~\cite{podili2017fast} and 1.55$\times$ better than~\cite{qiu2016going} but with a much higher throughput. 

By implementing conv. engine with parameters $m,r=4,3$, we achieved approximately 4.75$\times$ higher throughput while using only 2.67$\times$ more multipliers than~\cite{podili2017fast}. This is because of the larger output tile $m \times m=4 \times 4$ per PE per clock cycle and higher number of parallel PEs employed in our design. This increase in throughput is also reflected by our overall latency for the computation of VGG\,16 D. Our system takes only 28.05\,ms to compute the results of all 5 group layers of VGG\,16 D compared to 133.22\,ms and 163.40\,ms taken by~\cite{podili2017fast} and~\cite{qiu2016going}, respectively. Relative to older implementation~\cite{qiu2016going}, we achieved 5.83$\times$ throughput while using 0.88$\times$ multipliers. Using $m=4$, we achieved a multiplier efficiency of 1.60 GOPs/sec/multiplier in comparison to 0.90 and 0.24 GOPs/sec/multiplier for~\cite{podili2017fast} and~\cite{qiu2016going}, respectively.  

\begin{table}[tpb]
\caption{Performance comparison for VGG16 network D}
\vspace{-2mm}
\label{tab:Table2}
\begin{minipage}{\columnwidth}
\begin{center}
{
\begin{tabular}{|l@{\hspace{1mm}}|@{\hspace{1mm}}c@{\hspace{1mm}}|@{\hspace{1mm}}c@{\hspace{1mm}}|@{\hspace{1mm}}c@{\hspace{1mm}}|@{\hspace{1mm}}c@{\hspace{1mm}}|@{\hspace{1mm}}c@{\hspace{1mm}}|c|}
\hline 
       		
      & \cite{qiu2016going}  & \cite{podili2017fast}  & \cite{podili2017fast}\footnote{Normalized w.r.t. number of multipliers in our $F(2 \times 2,3 \times 3)$}   &\multicolumn{3}{c|}{Our proposed designs} \\ \hline
$m,r$	               & - &   2,3   & 2,3 &  2,3 & 3,3 & 4,3 \\ \hline
Number of multipliers      & 780 & 256 & 688 & 688 & 700 & 684 \\\hline
Number of PEs              & - & 16 & 43 & 43 & 28 & 19\\\hline
Data precision (bits)  & 16 & 32 & 32 & 32 & 32 & 32\\\hline
Frequency (MHz)        & 150 &  200 &200 &200 &200 &200 \\\hline
Conv1 (ms) & 31.29 & 16.81 & 6.25 & 6.25 & 4.27 & 3.54\\\hline
Conv2 (ms) & 23.58 & 24.08 & 8.96 & 8.96 & 6.12 & 5.07\\\hline
Conv3 (ms) & 39.29 & 40.14 & 14.94 & 14.94 & 10.19 & 8.45\\\hline
Conv4 (ms) & 36.30 & 40.14 & 14.94 & 14.94 & 10.19 & 8.45\\\hline
Conv5 (ms) & 32.95 & 12.04 & 4.48 & 4.48 & 3.06 & 2.54\\\hline
Overall latency (ms) & 163.4 & 133.22 & 49.57 & 49.57 & 33.83 & 28.05\\\hline
Throughput  & 187.8 & 230.4 & 619.2 & 619.2 & 907.2 & 1094.3\\
(GOPS/sec)  &  &  &  &  &  & \\\hline
Multiplier efficiency & 0.24 & 0.90 & 0.90 & 0.90 & 1.29 & 1.60\\
(GOPS/sec/multiplier)&  &  &  &  &  & \\\hline
Power (W) & 9.63 & 8.04 & 21.61 & 13.03 & 23.96 & 36.32\\\hline
Power efficiency  & 19.50 & 28.66 & 28.66 & 41.34 & 37.87 & 30.13\\
(GOPS/sec/W) &  &  &  &  &  & \\\hline

\hline
\end{tabular}
}
\end{center}
\vspace{-5mm}
\end{minipage}
\vspace{-3mm}
\end{table}

\section{Conclusions}
Convolutional Neural Networks (CNNs) have gained widespread popularity in the field of computer vision and image processing. Due to their huge computational demands, dedicated hardware e.g. FPGA-based implementations are being explored to improve the performance of CNNs. In this paper, we explored the design space associated with Winograd minimal filtering to find the sets of parameters that result in improved throughput and power-efficiency of CNNs. We also designed a pipelined and parallel Winograd convolution engine that improves the throughput and power-efficiency while reducing the computation complexity of the overall system. Experimental results show that our proposed designs provide up to 4.75$\times$ and 1.44$\times$ improvements in throughput and power-efficiency, respectively, in comparison to the state-of-the-art implementation while using 2.67$\times$ more multipliers. Furthermore, our design obtained logic resource savings of about 53.6\% to achieve better power-efficiency.

\section*{Acknowledgments}
\vspace{-1mm}

The authors would like to thank Andrew Lavin and Abhinav Podili for their continuous help and support in this work.

\bibliography{references}
\bibliographystyle{IEEEtran}

\end{document}